\documentclass[aps,superscriptaddress,twocolumn,showpacs,preprintnumbers,amsmath,amssymb]
{revtex4}
\usepackage{ae,graphicx,psfrag}

\newcommand{\be}{\begin{equation}}
\newcommand{\ee}{\end{equation}}

\begin{document}

\title{Non-universal behavior for aperiodic interactions within a mean-field
approximation}

\author{Maicon S. Faria}
\affiliation{Departamento de Física,
Universidade Federal de Santa Catarina, 
88040-900, Florianópolis, SC, Brazil}
\affiliation{Instituto de Física,
Universidade de São Paulo, Caixa Postal 66318,
05315-970, São Paulo, SP, Brazil }
\author{N. S. Branco}\email{nsbranco@fisica.ufsc.br}
\affiliation{Departamento de Física,
Universidade Federal de Santa Catarina, 
88040-900, Florianópolis, SC, Brazil} 
\author{M. H. R. Tragtenberg}
\affiliation{Departamento de Física,
Universidade Federal de Santa Catarina, 
88040-900, Florianópolis, SC, Brazil}

\date{\today}

\begin{abstract}

         We study the spin-$1/2$ Ising model on a Bethe
lattice in the mean-field limit, with the interaction constants following 
two deterministic aperiodic sequences: Fibonacci or period-doubling ones. 
New algorithms of sequence generation were implemented, which were fundamental
in obtaining long sequences and, therefore, precise results. 
We calculate the exact critical temperature for both sequences, as well as the
critical exponent $\beta$, $\gamma$ and $\delta$.
For the Fibonacci sequence, the exponents are classical,
while for the period-doubling one they depend on the ratio between the two
exchange constants. The usual relations
between critical exponents are satisfied, within error bars, for the
period-doubling sequence. Therefore, we show 
that mean-field-like procedures may lead to nonclassical critical
exponents.

\noindent     

\end{abstract}

\maketitle

\newpage

		Considerable attention has been devoted to the
investigation of systems displaying inhomogeneous or disordered interactions
\cite{cardy}. These
systems have experimental relevance, since many of the materials found in
nature come with impurities. Moreover, modern techniques are able to build
materials with controlled composition, such that two or more different atoms are
combined in a given order \cite{quasicristais}. Theoretically, one possible 
question is to which
extent the introduction of quenched disorder may affect the critical behavior
of an originally pure system \cite{cardy}. The answer to this question for random
disorder is given by the Harris
criterion \cite{harris}: if the pure-system's specific-heat exponent, $\alpha$, is
positive (negative), the critical behavior of the disordered model is diffferent 
from (the same as for) the pure model.

	The discovery of quasi-crystals \cite{quasicristais} has motivated 
the introduction of another kind of inhomogeneity in a system: 
its interactions are now modulated by aperiodic sequences, obtained from, for example,
deterministc substitution rules. These sequences have been a subject of
intense research in recent years \cite{seqs_aper}: numerical \cite{bercher} as well 
as analytical results \cite{vieira1} have been obtained and
the so-called Harris-Luck criterion is avalailable \cite{luck1}. 
According to this criterion, the relevance of a given
aperiodic sequence is connected to the crossover exponent 
$\phi = 1 - d_a \nu_0 \left( 1 - \omega \right) $, 
where $d_a$ is the dimension in which the aperiodic sequence acts, $\nu_0$ is the
correlation lenght's
critical exponent of the pure model and $\omega$ is defined through
$g \sim N^{\omega}$. Here, $g$ is the fluctuation in the number of a given letter of 
the sequence and $N$ is the length of the sequence (see below). Generally speaking, the greater 
the fluctuation of the
sequence the easier to move the critical behavior of the system
away from the homogeneous case. 
We can use many approaches to address the influence of aperiodicity
on critical behavior. Mean field is perhaps the most used approximation
in condensed matter
problems, ranging from complex fluids \cite{fluidoscomplexos} and superconductivity
\cite{supercondutividade} to Bose-Einstein condensation \cite{bec}
and magnetism \cite{leite,hoston}. One possible realization of a mean-field picture 
is the Bethe lattice: the critical exponents for homogeneous models 
in this geometry are classical (as
for usual mean-field approximations), namely: $\beta=1/2$, $\gamma=1$, and $\delta=3$,
for example. In the present case, we
arrive at the surprising result that aperiodic modulation of exchange constants
may change the critical exponent of a magnetic system \textit{within a mean-field
approximation}. Therefore, the application of this approach seems to be generally
useful in the study of critical phenomena in complex systems.

\begin{figure} %Figure1%
\includegraphics{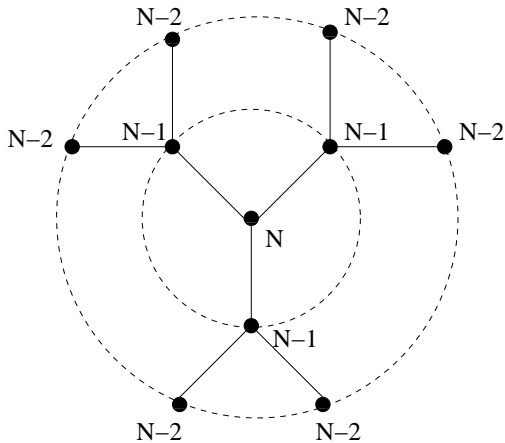}
\caption{\label{fig:bethe} Portion of a Bethe lattice with coordination number
$z=3$. Spins are represented by black dots. Generations are linked by dashed 
lines (which do \textbf{not} represent
interactions between spins) and the interaction constants between different
generations (continuous lines) may have two different values, $J_{A}$ or $J_{B}$.}
\end{figure}

	Our goal in this paper is twofold: investigate whether there are changes in the
universality class of magnetic systems with aperiodic interactions in high-dimensional
systems and determine the existence or not of nonclassical critical exponents within a
mean-field framework.
To achieve this, we define the Ising model on a Bethe lattice, which
has proven to be a good approximation to the critical behavior in dimension 
three or above. In this geometry,
no closed loops are allowed: see Fig.~\ref{fig:bethe} for
an example of a portion of a Bethe lattice with $z=3$.
The models we study is defined by the Hamiltonians:
\be  {\cal H}_{1/2} = - \sum_{<i,j>} J_{n} s_i s_j, \;\;\; s_i=\pm 1,
\label{eq:hamil_ising} \ee
where the sum is over nearest-neighbor spins and
$J_{n}$ is the exchange constant between
sites on generations $n$ and $n+1$ and may have different values, depending 
on the generations and the corresponding 
letter in the aperiodic sequence (although all
interactions between the same pair of generations have the same value).

	The exchange constant $J_{n}$ is chosen
according to the respective letter in the aperiodic sequence. The construction of the
aperiodic sequence is made in the Bethe lattice from the exterior to the 
interior, i.e, the first
letters of the sequence correspond to the interactions between generations in
the exterior of the lattice. The sequences we have studied are the following:
\begin{itemize}
	\item[(i)] Fibonacci sequence, built from the substitution rules: 
$A \rightarrow AB, 
B \rightarrow A$. The first stages of this sequence are: 
$A \rightarrow AB \rightarrow ABA \rightarrow
ABAAB \rightarrow ABAABABA $. This last finite sequence corresponds to the following
sequence of interaction constants: $J_A J_B J_A J_A J_B J_A J_B J_A$;
	\item[(ii)] period-doubling sequence: in this case the substitution rules are 
$A \rightarrow AB$ and $B \rightarrow AA$. The first letters of the interaction
constant sequence are: $J_A J_B J_A J_A J_A J_B J_A J_B$.
\end{itemize}

	The geometrical characteristics of aperiodic sequences can be obtained
from the substitution matrix ${\cal M}$, which connects the number of letters
$A$ and $B$ after one application of the iteration rule, namely:
\be
        \left( \begin{array}{c} 
               N_{A}^{(n+1)} \\
               N_{B}^{(n+1)} 
               \end{array} \right) 
               = {\cal M}
        \left( \begin{array}{c}
               N_{A}^{(n)} \\
               N_{B}^{(n)} \end{array} \right),   \label{eq:matrix}
\ee
where $N_{\zeta}^{(n)}$ is the number of letters $\zeta$ ($A$ or $B$) after
$n$ iterations of the substitution rule. In particular, the total number of 
letters, $N^{(n)}$ grows exponentially with the number of iterations $n$:
\be
 N^{(n)} \sim \lambda_1^n, \; \mbox{as} n\rightarrow\infty, \label{eq:Ncomn}  
\ee
where $\lambda_1$  is the greater
eigenvalue of ${\cal M}$. The fraction of letters $A$ ($B$), $p_{A}$
($p_{B}$) in the infinite word, i.e,  after $n$ applications of the iteration rules,
for large $n$, is proportional to the first (second) entry of the eigenvector
corresponding to the greater eigenvalue of ${\cal M}$ (see Ref. \onlinecite{branco4}
for more details).
The fluctuation $g^{(n)}$ of a given letter, say $A$, is defined as:
\begin{equation}
   g^{(n)} = N_{A}^{(n)} - p_{A} N^{(n)}. \label{eq:fluctuation}
\end{equation}
It is possible to show that $g \sim N^{\omega}$, where $\omega=\ln |\lambda_2| / \ln \lambda_1$
($\lambda_2$ is the smaller eigenvalue of the substitution matrix). In a linear chain,
the fluctuation is greater for the period-doubling sequence than for the
Fibonacci one (see below for a discussion of this point).

	The solution on a Bethe lattice is obtained from a mapping between partial
magnetizations
of two consecutive generations; see \cite{salinas} for a detailed calculation 
of this mapping. In the mean-field limit,
$z\rightarrow \infty, J_{n} \rightarrow 0$, such that the product 
$\widetilde{J}_n \equiv zJ_{n}$ is finite, and for external magnetic field $h$,
the expression
simplifies a great deal and we obtain:
\be  m_{n+1} =  \tanh ( \widetilde{K}_n m_{n} + H),  \label{eq:mapaIsing}  \ee
where $m_i$ is the partial magnetization in generation $i$, $\widetilde{K}_n \equiv 
\widetilde{J}_n/k_B T$, $T$ is the temperature, $k_B$ is the Boltzmann constant,
and $H = h/k_B T$. 
Note that we took the mean-field limit in order to simplify the equations and calculate
the critical temperature analitically.

	The critical points of the model studied here are 
at zero external field. The exact critical temperature is given
by the stability limit of the paramagnetic phase, obtained from:
\be \left. \frac{d m_n}{d m_1}\right|_{\{m\}=0} = 1, \label{eq:deriTc} \ee
where $m_n$ is the magnetization of spins in generation $n (n \rightarrow \infty)$, 
$m_1$ is the magnetization
of spins in generation $1$ and $\{m\}=0$ means that the derivatives are 
taken in the point 
where the magnetization of all generations from $1$ to $n-1$ are zero. We
then obtain:
\be  \frac{k_B T_c}{\tilde{J}_A} = (1+r)^{p_B} \label{eq:Tcising},   \ee
where $T_c$ is the critical temperature, 
$(1+r) \equiv \tilde{J}_B / \tilde{J}_A$,  
and $p_B$ is the fraction of interactions $J_B$ in the thermodynamic limit. 
As discussed above, this fraction is proportional  to the second
entry of the eigenvector corresponding to the largest eigenvalue of the matrix 
${\cal M}$ \cite{branco4}. It is important
to know $T_c$ exactly (or with a very good precision) in order to obtain 
reliable values for the exponent $\beta$. 

	The magnetization is not uniform; rather, it follows an aperiodic sequence.
Therefore,
we will calculate and discuss the behavior of the mean magnetization, defined by:
\be \overline{m} \equiv \frac{1}{N} \sum_{i=1}^{N} m_i, \label{eq:magmedia} \ee
where $m_i$ is the partial magnetization in generation $i$, calculated 
through Eq.~(\ref{eq:mapaIsing}). This
is also the physical quantity which would be acessible to experiments.
In order to calculate $\gamma$, we define the zero-field
susceptibility $\chi = \partial \overline{m} / \partial h$, related
to the mean magnetization.

\begin{figure} %Figure2%
\includegraphics[scale=0.25]{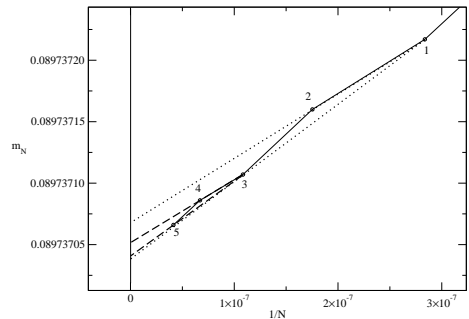}
\caption{\label{fig:extrapolacao} Magnetization as a function of 
the number of generations of
the Bethe lattice or, equivalently, of the number of letters of the aperiodic 
sequence, for the
Fibonacci sequence. We show a linear extrapolation procedure to obtain the 
thermodynamic
limit for the magnetization and its error, for $r=4$ (see text),
$102,334,155$ generations and $\log_{10}(t) = 10^{-2.5}$.}
\end{figure}

	At this point, it is worth recalling that for systems such
that the length-scaling parameter, in the renormalization-group context, 
\textit{cannot} assume
any value, thermodynamic quantities may follow log-periodic scaling laws 
\cite{niemeijer,andrade1,karevski}. The
mean magnetization at zero external magnetic field, e.g., is written as:
\be \overline{m} = t^{\beta} P\left[ \log (t) \right], \label{eq:maglogper} \ee
where $t=(T_c-T)/T_c$ is the reduced temperature and $P[x]$ is a 
periodic function of $x$.
Therefore, the logarithmic derivative of $\overline{m}$ at $h=0$ is given by:
\be 
\frac{d \log \overline{m}}{d \log t} = \beta + \widetilde{P}\left[ \log (t) \right],
\label{eq:derivada1} 
\ee
where $\widetilde{P}[x]$ is also a periodic function of $x$. In an analogous way:
\be 
\frac{d \log \overline{m}}{d \log h} = 1/\delta + \widetilde{Q}\left[ \log (h) \right],
\;\;\;  T=T_c,
\label{eq:derivada2} 
\ee
and
\be 
\frac{d \log \chi}{d \log t} = \gamma + \widetilde{R}\left[ \log (t) \right],
\;\;\; h=0,
\label{eq:derivada3} 
\ee
where $\widetilde{Q}$ and $\widetilde{R}$ are periodic functions of their
arguments. Our strategy is to calculate the derivatives,
do a best fit to a log-periodic function plus a constant and obtain the value of 
the respective exponents from this fit. Some points are
worth stressing: in order to calculate this derivative numerically, we need two values 
of $\overline{m}$ 
\textit{in the thermodynamic limit}. This is done in the following way. We start 
with an arbitrary
value of $m_1$ and iterate Eq.~(\ref{eq:mapaIsing}) many times, 
keeping track of the mean
magnetization along the process (the number of
iterations we used to obtain these results are about $10^{8}$ for the Fibonacci
sequence and $\sim 10^{11}$ for the period-doubling sequence).
After a transient and a sufficient number of iterations, 
the mean magnetization fluctuates around a monotonic \textit{trend} 
(see Fig. \ref{fig:extrapolacao}),
which does not depend on the initial value $m_1$. From these values,
we can extrapolate to the thermodynamic limit. This procedure can be done
for zero or non-zero magnetic field; therefore, we can calculate 
$\overline{m}$ as function of $t$ or $h$ and the susceptibility $\chi$.

In Fig. \ref{fig:extrapolacao} we depict $m_N \time 1/N$ for the Fibonacci
sequence for $h=0$, where $N$ is the total number of generations of the Bethe lattice or,
equivalently, the number of letters generated for the sequence, together
with the extrapolation procedure. We take two consecutive pairs of values
of $m_N$ and make a linear extrapolation of each pair to the limit
$1/N \rightarrow \infty$. The extrapolated value is taken as the mean of the
two values obtained and its error is estimated as half the interval. This
is an overestimation of the error but we are sure the true extrapolated value is within the interval.
We have compared our procedure with more formal extrapolations
like the so-called VBS \cite{vbs} and BST \cite{bst}. The extrapolated 
value is the same, although
the error estimation is somewhat arbitrary for the VBS algorithm and
clearly underestimated for the
BST one. With this procedure, we were able to calculate the derivative in 
Eqs. (\ref{eq:derivada1}), (\ref{eq:derivada2}) and (\ref{eq:derivada3}) in 
the thermodynamic limit. 

	A crucial point should be mentioned here: only using new algorithms were we
able to generate very long sequences for the period-doubling
case. With these algorithms, we need to store only $N$ letters, in order to
use an $N^2$-letter sequence. Long sequences allow for a reliable and precise extrapolation
to the thermodynamic limit (see below). The algorithm, for the period-doubling
sequence, works as follows. If an $N^2$-letter sequence, with $N=2^n$, $n$
integer, is generated and it is written in lines of length $N$,
one over the other,
it is easy to note that the first $N-1$ colums have the same letter as in the first line
of the respective column,
while the last column repeats the first line if $n$ is even or
repeat the first line with $A$ and $B$ swapped over if $n$ is odd.
Therefore, one needs to generate and store only the
first line, with $N$ letters, to be able to work with an $N^2$ sequence. This
algorithm, although with more complicated rules, can be used for other sequences,
which allow for economy in time and space.

\begin{figure} %Figure3%
\includegraphics[scale=0.25]{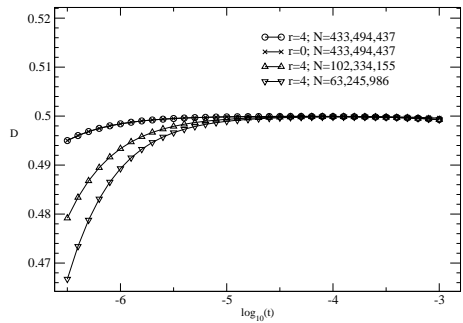}
\caption{\label{fig:derxtFibo} Log-derivative of the magnetization as a function of 
the reduced temperature
for the Fibonacci sequence. We show the behavior for different values
of $r$ and for different number of generations of the Bethe lattice. For comparison, the
result for the uniform model ($r=0$) is also depicted. It is clear that $\beta$
is the same for the uniform and aperiodic models.}
\end{figure}

\begin{figure} %Figure4%
\includegraphics[scale=0.25]{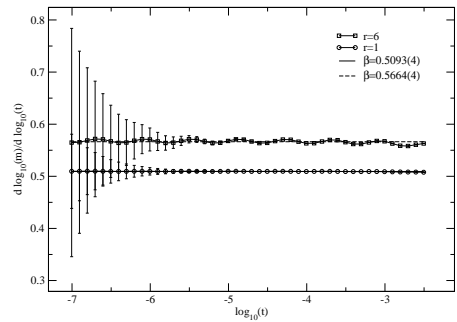}
\caption{\label{fig:derxtDupli} Log-derivative of the magnetization as a function 
of the reduced temperature for the period-doubling sequence, with $r=1$ and $r=6$.
The log-periodic
behavior is clearly identified and the values for $\beta$ are different from the 
classical  value $1/2$. The continuous line and circles (traced line and squares)
represent $r=1$ ($r=6$) and the length of the sequences is $2^{32}$.}
\end{figure}

\begin{figure} %Figure7%
\includegraphics[scale=0.25]{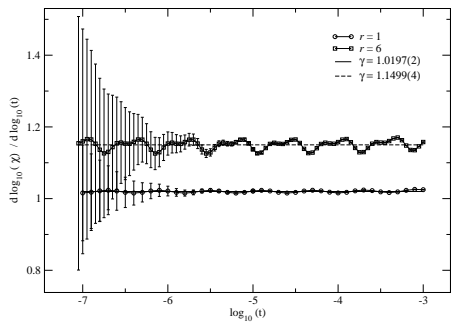}
\caption{\label{fig:derxtDupligamma} Log-derivative of the susceptibility as a function 
of the reduced temperature for the period-doubling sequence, with $r=1$
and $6$. The exponent $\gamma$ is different from the 
classical  value $1$ and depends on $r$
The continuous line and circles (traced line and squares)
represent $r=1$ ($r=6$) and the length of the sequences is $2^{32}$.}
\end{figure}

	Our results can be summarized in the following figures, where we depict 
the derivatives in Eqs. (\ref{eq:derivada1}), (\ref{eq:derivada2}) and (\ref{eq:derivada3}).
These equations tell us that the log-derivatives are represented by
periodic functions of
$\log (t)$ or $\log h$, with a mean given by the value of the respective exponent. For the Fibonacci
sequence (Fig.~\ref{fig:derxtFibo}), it is clear that the aperiodic modulation does not
change the  exponent $\beta$ from its classical value, $1/2$. 
Note the behavior of the log-derivative when the size
of the Bethe lattice (or, equivalently, of the aperiodic sequence) is 
increased: for large values
of the reduced temperature $t$, we are outside of the scaling region and Eq.~(\ref{eq:derivada1})
is not expected to be obeyed. When
the reduced temperature is decreased, the correlation length increases and eventually it
grows bigger than the number of generations of the Bethe lattice, 
leading to the finite-size effect seen in the left-hand size of the
curves (they decrease from the classical value 1/2).  When the size 
of the lattice is increased,
this departure from the infinite-size behavior takes place at lower values of $t$,
as expected. This finite-size effect is also represented by the size of the error
bars, as depicted in Figs. \ref{fig:derxtDupli}, \ref{fig:derxtDuplidelta} and 
\ref{fig:derxtDupligamma} (we will comment more on this aspect below). The same overall behavior is obtained
for $\gamma$ and $\delta$, which assume their classical values, $1$ and $3$, respectively.
	
\begin{figure} %Figure5%
\includegraphics[scale=0.25]{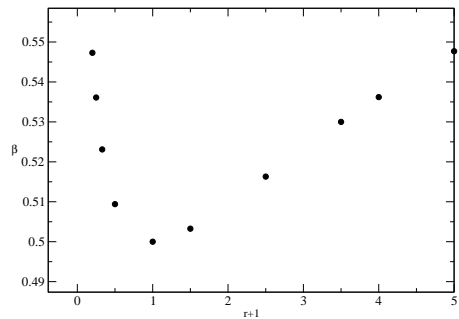}
\caption{\label{fig:betaxrho} Dependence of the exponent $\beta$ on the parameter $r$.
The homogeneous system is represented by $r=0$.}
\end{figure}

	The scenario is more interesting for the period-doubling sequence. 
Although the sequences we need to generate are bigger than 
in the Fibonacci case, we were able
to determine precise values for the exponents $\beta$, $\gamma$
and $\delta$, which 
\textit{depend on the interaction
ratio $r$}. In Fig.~\ref{fig:derxtDupli} we show the log-derivative of
$\overline{m}$ as a  function of $\log(t)$ for
$r=1$ and $r=6$. The scaling region spans three decades and the exponent is clearly 
different from the
classical value $1/2$: a fitting of the data to a log-periodic function leads to
$\beta=0.5093(4)$ and $\beta=0.5664(5)$ for $r=1$ and $r=6$, respectively.
The exponent $\beta$ depends on the ratio $r$, as 
depicted in Fig.~\ref{fig:betaxrho}:
the overall behavior is the same as the one found in Ref.~\onlinecite{igloi}, but 
for a different sequence. There is a symmetry with respect to $r=0$: the exponent $\beta$
is the same for $1+r=l$ and $1+r=1/l$, i.e, it is invariant with respect to the exchange
$J_A \leftrightarrow J_B$. The exponent $\beta$ increases for the
aperiodic model, when compared to its uniform value. This is consistent with
the following picture: the new (aperiodic) fixed points, in the renormalization-group sense,
must have a greater value of $\nu$ than the one for the uniform model ($\nu_0$). Therefore,
assuming $\nu=1/y_t$ to be valid, where $y_t$ is the usual temperature scaling exponent 
in the renormalization-group equations, $y_t$ is smaller for the aperiodic model. But
$\beta=(d-y_h)/y_t$ ($y_h$ is the field scaling exponent) and $y_h$ is seen to vary very 
little with $r$. In fact, our results for $\delta$ (see Fig. \ref{fig:derxtDuplidelta}), 
given by $y_h/(d-y_h)$,  show that, although this exponent
depends on $r$ for the period-doubling sequence, it varies much less than
$\beta$. So, the variation of this exponent is linked almost entirely to $y_t$ and
it increases for the aperiodic model (in fact, this picture is also observed for random
disorder). In  Fig. \ref{fig:derxtDupligamma}, we
depict $d (\log \chi) / d (\log t)$ as a function of $\log t$, for $r=1$ and $6$:
again, it is evident the dependence on $r$. It is clear, from this figure,
the need of a second harmonic, to adjust the curve for $r=6$. The increase 
of the error bars for small values of $\log (t)$ in Figs.
\ref{fig:derxtDupli}, \ref{fig:derxtDuplidelta} and \ref{fig:derxtDupligamma}, 
is a finite-size effect: the smaller $t$, the greater the correlation length, and
the size of the Bethe lattice (or, equivalently, the size of the aperiodic
sequence) must be increased, to reach the limit where the infinite-volume
behavior is obtained.

We summarize our results in Table
\ref{tab:exponents}, where we show the values of the three exponents for the
period-doubling sequence and $r=0$, $1$ and $6$. For $r=0$ we show the exact results
on the Bethe lattice: our numerical values agree with them within erros bars. The
exponents clearly depend on $r$;
nevertheless, a usual relation among the three exponents, namely 
$\beta=\gamma/(\delta-1)$ is satisfied for $r=1$ and is just satisfied
for $r=6$ (see lines 3 and 4 in Table \ref{tab:exponents}). 

\begin{figure} %Figure6%
\includegraphics[scale=0.25]{deltaxloghG36r1_6}
\caption{\label{fig:derxtDuplidelta} Log-derivative of the magnetization as a function 
of the magnetic field, at $T=T_c$, for the period-doubling sequence, with $r=1$
and $6$. The exponent $1/\delta$ is different from the 
classical  value $1/3$ and depends on $r$
 The continuous line and circles (traced line and squares)
represent $r=1$ ($r=6$) and the length of the sequences is $2^{36}$.}
\end{figure}

	Preliminary results obtained for the
spin-$1$ Ising model lead to exactly the same behavior as for its
spin-$1/2$ counterpart. As the former model is the Blume-Capel (BC) one with
zero crystal field $\Delta$ and the latter corresponds to the BC model
with $\Delta=-\infty$, we can infer that our results hold for the BC Hamiltonian
and for all values of $\Delta$ such that the transition is continuous in the 
uniform case. Also, the results we obtain for the
Blume-Capel model are not expected to depend on which interaction the disorder acts on,
according to the following renormalization-group reasoning. Even if, initially, only
the exchange constants follow an aperiodic sequence, the crystal field will not
be uniform in the coarse-grained system, which is equivalent to the original one.

	As a final note, we would like to mention that Eq. (\ref{eq:matrix})  assumes
that the relation between the number of letters A or B on stage $n+1$ of the iteration process
depends linearly on the number of letters A and B on stage $n$, namely: 
\be
        \begin{array}{ccc} 
               N_A^{(n+1)} & = & a N_A^{(n)} + b N_B^{(n)} \\
               N_B^{(n+1)} & = & c N_A^{(n)} + d N_B^{(n)}, 
               \end{array}    \label{eq:linear}
\ee
where $a$, $b$, $c$, and $d$ are the elements ot the matriz ${\cal M}$. The former 
relation does not hold true for the Bethe lattice, since letters $A$ on different 
generations of the lattice do not lead to the same number of letters $A$ and $B$ 
(the same is true for the substitution of letters $B$) on the inflation process. Then, as far as we
cannot define the substitution matrix for aperiodic sequences on Bethe lattices,
we are not able to determine the exponent $\omega$ in these cases. 

	In summary, we have studied the Ising model on
a Bethe lattice, in the mean-field limit, such that the exchange constants follow
two different aperiodic sequences: the Fibonacci and period-doubling ones.
The results are as follows. For the Fibonacci sequence,
the exponent are classical, namely $\beta=1/2$, $\gamma=1$ and $\delta=3$. For the 
period-doubling sequence the exponents depend on the ratio of the interaction constants,
but satisfying a usual relation among critical exponents for any value of $r$.
Therefore, we show  that mean-field-like procedures may lead to
exponents' values different from the classical ones. In this particular  case, although
thermal fluctuations are supressed by the mean-field character of the procedure we 
use, geometrical fluctuations present in the sequences are treated exactly. 
Preliminary results for lattices with finite coordination number $z$ show the same 
qualitative behavior as for the mean-field limit ($z \rightarrow \infty$).

\begin{table}
\caption{\label{tab:exponents} Values for the exponents
$\beta$, $\gamma$ and $\delta$ for the period-doubling sequence,
for $r=0$, $1$ and $6$. In the third and fourth lines we show evaluations
of $\beta$, calculated from $\gamma$ and $\delta$ and
directly from our data, respectively. Note that the results in the first column are
exact ones: our numerical values agree with them, within error bars. Numbers
inside parenthesis are errors in the last decimal figures.}
\begin{ruledtabular}
 \begin{tabular}{|c|c|c|c|} 
                          & $r=0$ &  $r=1$     &  $r=6$ \\ \hline 
   $\gamma$               &   1   &  1.0197(2) & 1.1499(4) \\ \hline
   $\delta$               &   3   &  3.0006(2) & 3.0266(9) \\ \hline
$\frac{\gamma}{\delta-1}$ &  1/2  &  0.5097(2) & 0.5674(5) \\ \hline
   $\beta$                &  1/2  &  0.5093(4) & 0.5664(5) \\
 \end{tabular}
\end{ruledtabular}
\end{table}

\begin{acknowledgments}
The authors would like to thank Dr. W. Figueiredo, Dr. J. A. Plascak,
and Dr. Andr\'e P. Vieira for interesting
discussions and FAPESC, CNPq, and CAPES for partial financial support.
We thank S. R. Salinas for a careful reading of the manuscript and for
many helpful sugestions.
\end{acknowledgments}

\bibliography{references}

\begin{thebibliography}{20}
\expandafter\ifx\csname natexlab\endcsname\relax\def\natexlab#1{#1}\fi
\expandafter\ifx\csname bibnamefont\endcsname\relax
  \def\bibnamefont#1{#1}\fi
\expandafter\ifx\csname bibfnamefont\endcsname\relax
  \def\bibfnamefont#1{#1}\fi
\expandafter\ifx\csname citenamefont\endcsname\relax
  \def\citenamefont#1{#1}\fi
\expandafter\ifx\csname url\endcsname\relax
  \def\url#1{\texttt{#1}}\fi
\expandafter\ifx\csname urlprefix\endcsname\relax\def\urlprefix{URL }\fi
\providecommand{\bibinfo}[2]{#2}
\providecommand{\eprint}[2][]{\url{#2}}

\bibitem[{\citenamefont{Cardy}(1996)}]{cardy}
\bibinfo{author}{\bibfnamefont{J.}~\bibnamefont{Cardy}},
  \emph{\bibinfo{title}{Scaling and Renormalization in Statistical Physics}}
  (\bibinfo{publisher}{Cambridge University Press, Cambridge, UK},
  \bibinfo{year}{1996}).

\bibitem[{\citenamefont{Shechtman et~al.}(1984)\citenamefont{Shechtman, Blech,
  Gratias, and Cahn}}]{quasicristais}
\bibinfo{author}{\bibfnamefont{D.}~\bibnamefont{Shechtman}},
  \bibinfo{author}{\bibfnamefont{I.}~\bibnamefont{Blech}},
  \bibinfo{author}{\bibfnamefont{D.}~\bibnamefont{Gratias}}, \bibnamefont{and}
  \bibinfo{author}{\bibfnamefont{J.~W.} \bibnamefont{Cahn}},
  \bibinfo{journal}{Phys. Rev. Lett.} \textbf{\bibinfo{volume}{53}},
  \bibinfo{pages}{1951} (\bibinfo{year}{1984}).

\bibitem[{\citenamefont{Harris}(1974)}]{harris}
\bibinfo{author}{\bibfnamefont{A.~B.} \bibnamefont{Harris}},
  \bibinfo{journal}{J. Phys. C: Sol. St. Phys.} \textbf{\bibinfo{volume}{7}},
  \bibinfo{pages}{1671} (\bibinfo{year}{1974}).

\bibitem[{\citenamefont{Grimm and Baake}(1997)}]{seqs_aper}
\bibinfo{author}{\bibfnamefont{U.}~\bibnamefont{Grimm}} \bibnamefont{and}
  \bibinfo{author}{\bibfnamefont{M.}~\bibnamefont{Baake}}, in
  \emph{\bibinfo{booktitle}{The Mathematics of Long-Range Aperiodic Order}},
  edited by \bibinfo{editor}{\bibfnamefont{R.~V.} \bibnamefont{Moody}}
  (\bibinfo{publisher}{Kluwer}, \bibinfo{year}{1997}), pp.
  \bibinfo{pages}{199--237}.

\bibitem[{\citenamefont{Berche et~al.}(1998)\citenamefont{Berche, Chatelain,
  and Berche}}]{bercher}
\bibinfo{author}{\bibfnamefont{P.~E.} \bibnamefont{Berche}},
  \bibinfo{author}{\bibfnamefont{C.}~\bibnamefont{Chatelain}},
  \bibnamefont{and} \bibinfo{author}{\bibfnamefont{B.}~\bibnamefont{Berche}},
  \bibinfo{journal}{Phys. Rev. Lett.} \textbf{\bibinfo{volume}{80}},
  \bibinfo{pages}{297} (\bibinfo{year}{1998}).

\bibitem[{\citenamefont{Vieira}(2005)}]{vieira1}
\bibinfo{author}{\bibfnamefont{A.~P.} \bibnamefont{Vieira}},
  \bibinfo{journal}{Phys. Rev. Lett.} \textbf{\bibinfo{volume}{94}},
  \bibinfo{pages}{077201} (\bibinfo{year}{2005}).

\bibitem[{\citenamefont{Luck}(1993)}]{luck1}
\bibinfo{author}{\bibfnamefont{J.~M.} \bibnamefont{Luck}},
  \bibinfo{journal}{Europhys. Lett.} \textbf{\bibinfo{volume}{24}},
  \bibinfo{pages}{359} (\bibinfo{year}{1993}).

\bibitem[{\citenamefont{Lagerwall and Giesselmann}(2006)}]{fluidoscomplexos}
\bibinfo{author}{\bibfnamefont{J.~P.~F.} \bibnamefont{Lagerwall}}
  \bibnamefont{and}
  \bibinfo{author}{\bibfnamefont{F.}~\bibnamefont{Giesselmann}},
  \bibinfo{journal}{Chem. Phys. Chem.} \textbf{\bibinfo{volume}{7}},
  \bibinfo{pages}{20} (\bibinfo{year}{2006}), \bibinfo{note}{\uppercase{J}.
  Galanis, D. Harries, D. L. Sackett, W. Losert, and R. Nossal, Phys. Rev.
  Lett. {\bf 96}, 028002 (2006)}.

\bibitem[{\citenamefont{P.~A.~Lee and Wem}(2006)}]{supercondutividade}
\bibinfo{author}{\bibfnamefont{N.~N.} \bibnamefont{P.~A.~Lee}}
  \bibnamefont{and} \bibinfo{author}{\bibfnamefont{X.-G.} \bibnamefont{Wem}},
  \bibinfo{journal}{Rev. Mod. Phys.} \textbf{\bibinfo{volume}{78}},
  \bibinfo{pages}{17} (\bibinfo{year}{2006}), \bibinfo{note}{\uppercase{B}.
  Kyung and A. M. S. Tremblay, Phys. Rev. Lett. {\bf 97}, 046402 (2006)}.

\bibitem[{\citenamefont{Davis and Blakie}(2006)}]{bec}
\bibinfo{author}{\bibfnamefont{M.~J.} \bibnamefont{Davis}} \bibnamefont{and}
  \bibinfo{author}{\bibfnamefont{P.~B.} \bibnamefont{Blakie}},
  \bibinfo{journal}{Phys. Rev. Lett.} \textbf{\bibinfo{volume}{96}},
  \bibinfo{pages}{060404} (\bibinfo{year}{2006}), \bibinfo{note}{\uppercase{O}.
  Morsch and M. Oberthaler, Rev. Mod. Phys. {\bf 78}, 179 (2006)}.

\bibitem[{\citenamefont{Leite and Figueiredo}(2006)}]{leite}
\bibinfo{author}{\bibfnamefont{V.~S.} \bibnamefont{Leite}} \bibnamefont{and}
  \bibinfo{author}{\bibfnamefont{W.}~\bibnamefont{Figueiredo}},
  \bibinfo{journal}{Phys. Rev. B} \textbf{\bibinfo{volume}{74}},
  \bibinfo{pages}{094408} (\bibinfo{year}{2006}), \bibinfo{note}{and references
  therein}.

\bibitem[{\citenamefont{Hoston and Berker}(1991)}]{hoston}
\bibinfo{author}{\bibfnamefont{H.}~\bibnamefont{Hoston}} \bibnamefont{and}
  \bibinfo{author}{\bibfnamefont{A.~N.} \bibnamefont{Berker}},
  \bibinfo{journal}{Phys. Rev. Lett.} \textbf{\bibinfo{volume}{67}},
  \bibinfo{pages}{1027} (\bibinfo{year}{1991}).

\bibitem[{\citenamefont{Branco et~al.}(2007)\citenamefont{Branco, de~Sousa, and
  Ghosh}}]{branco4}
\bibinfo{author}{\bibfnamefont{N.~S.} \bibnamefont{Branco}},
  \bibinfo{author}{\bibfnamefont{J.~R.} \bibnamefont{de~Sousa}},
  \bibnamefont{and} \bibinfo{author}{\bibfnamefont{A.}~\bibnamefont{Ghosh}},
  \bibinfo{journal}{submitted to Phys. Rev. E}  (\bibinfo{year}{2007}).

\bibitem[{\citenamefont{Osório et~al.}(1989)\citenamefont{Osório, de~Oliveira,
  and Salinas}}]{salinas}
\bibinfo{author}{\bibfnamefont{R.}~\bibnamefont{Osório}},
  \bibinfo{author}{\bibfnamefont{M.~J.} \bibnamefont{de~Oliveira}},
  \bibnamefont{and} \bibinfo{author}{\bibfnamefont{S.~R.}
  \bibnamefont{Salinas}}, \bibinfo{journal}{J. Phys.: Condens. Matter}
  \textbf{\bibinfo{volume}{1}}, \bibinfo{pages}{6887} (\bibinfo{year}{1989}).

\bibitem[{\citenamefont{Niemeijer and van Leeuwen}(1976)}]{niemeijer}
\bibinfo{author}{\bibfnamefont{T.}~\bibnamefont{Niemeijer}} \bibnamefont{and}
  \bibinfo{author}{\bibfnamefont{J.~M.~J.} \bibnamefont{van Leeuwen}},
  \emph{\bibinfo{title}{Phase Transitions and Critical Phenomena}},
  vol.~\bibinfo{volume}{6} (\bibinfo{publisher}{Academic, New York, USA},
  \bibinfo{year}{1976}).

\bibitem[{\citenamefont{Andrade}(2000)}]{andrade1}
\bibinfo{author}{\bibfnamefont{R.~F.~S.} \bibnamefont{Andrade}},
  \bibinfo{journal}{Phys. Rev. E} \textbf{\bibinfo{volume}{61}},
  \bibinfo{pages}{7196} (\bibinfo{year}{2000}).

\bibitem[{\citenamefont{Karevski and Turban}(1996)}]{karevski}
\bibinfo{author}{\bibfnamefont{D.}~\bibnamefont{Karevski}} \bibnamefont{and}
  \bibinfo{author}{\bibfnamefont{L.}~\bibnamefont{Turban}},
  \bibinfo{journal}{J. Phys. A: Math. Gen.} \textbf{\bibinfo{volume}{29}},
  \bibinfo{pages}{3461} (\bibinfo{year}{1996}).

\bibitem[{\citenamefont{Hamer and Barber}(1981)}]{vbs}
\bibinfo{author}{\bibfnamefont{C.~J.} \bibnamefont{Hamer}} \bibnamefont{and}
  \bibinfo{author}{\bibfnamefont{M.~N.} \bibnamefont{Barber}},
  \bibinfo{journal}{J. Phys. A: Math. Gen.} \textbf{\bibinfo{volume}{14}},
  \bibinfo{pages}{2009} (\bibinfo{year}{1981}).

\bibitem[{\citenamefont{Henkel and Schütz}(1988)}]{bst}
\bibinfo{author}{\bibfnamefont{M.}~\bibnamefont{Henkel}} \bibnamefont{and}
  \bibinfo{author}{\bibfnamefont{G.}~\bibnamefont{Schütz}},
  \bibinfo{journal}{J. Phys. A: Math. Gen.} \textbf{\bibinfo{volume}{21}},
  \bibinfo{pages}{2617} (\bibinfo{year}{1988}).

\bibitem[{\citenamefont{Iglói and Palágyi}(1987)}]{igloi}
\bibinfo{author}{\bibfnamefont{F.}~\bibnamefont{Iglói}} \bibnamefont{and}
  \bibinfo{author}{\bibfnamefont{G.}~\bibnamefont{Palágyi}},
  \bibinfo{journal}{Physica A} \textbf{\bibinfo{volume}{240}},
  \bibinfo{pages}{685} (\bibinfo{year}{1987}).

\end{thebibliography}

\end{document}